\documentclass{aastex6}

\begin{document}


\title{Particles Co-orbital to Janus and Epimetheus: A Firefly Planetary Ring}

%% Use \author, \affil, plus the \and command to format author and affiliation 
%% information.  If done correctly the peer review system will be able to
%% automatically put the author and affiliation information from the manuscript
%% and save the corresponding autThe ring has also been identified at lower phase angles (~146 deg and ~157 deg) with normal I/F at least one order of magnitude lower than those presented in this work (Matthew Hedman, private communication). hor the trouble of entering it by hand.
%%
%% The \affil should be used to document primary affiliations and the
%% \altaffil should be used for secondary affiliations, titles, or email.

%% Authors with the same affiliation can be grouped in a single
\author{Othon C. Winter\altaffilmark{1}, Alexandre P.S. Souza, Rafael Sfair, Silvia M. Giuliatti Winter, Daniela C. Mour\~ ao}
\affil{S\~ ao Paulo State University - UNESP\\
Grupo de Din\^ amica Orbital \& Planetologia\\
Guaratinguet\' a, CEP 12516-410, SP, BRAZIL}

%\author{Butler Burton\altaffilmark{3}}
%\affil{National Radio Astronomy Observatory}

%\affil{TeXnology Inc}

%\author{Julie Steffen\altaffilmark{4}}
%\affil{American Astronomical Society \\
%2000 Florida Ave., NW, Suite 300 \\
%Washington, DC 20009-1231, USA}

%% Use the \and command so offset the last author.
%\and

\author{Dietmar W. Foryta}
\affil{Universidade Federal do Paran\' a - UFPr\\
Curitiba, PR, BRAZIL}


\altaffiltext{1}{ocwinter@gmail.com}

\begin{abstract}

 The Cassini spacecraft found a new and unique ring that shares the trajectory 
   of Janus and Epimetheus, co-orbital satellites of Saturn.
 Performing image analysis, we found this to be a continuous ring.
 Its width is between $30\%$ and $50\%$ larger than previously announced.
  We also verified that the ring behaves like a firefly. It can only 
  be seen from time to time, when Cassini, the ring and 
  the Sun are arranged in a particular geometric configuration, in very high phase angles. 
  Otherwise, it remains `in the dark', not visible to Cassini's cameras. 
  Through numerical simulations, we found a very short lifetime for the ring particles, less than
  a couple of decades. Consequently, the ring needs to be constantly 
  replenished. 
  Using a model of particles production due to micrometeorites impacts on the surfaces 
  of Janus and Epimetheus, we reproduce the ring, explaining its existence and the `firefly' behavior.

%Using a simple collisional model we verified that this phenomenon 
%is a natural outcome of the size distribution of the ring particles and also its relatively small amount of particles.

\end{abstract}

\keywords{planets, satellites: individual (Epimetheus, Janus) -– planets and satellites: dynamical evolution and
stability -– planets and satellites: rings}



\section{Introduction} \label{sec:intro}

Among its early findings, the Cassini Imaging Team discovered a ring  occupying the region of the trajectories of
 the satellites Janus and Epimetheus \citep{b1}.
These two Saturnian satellites are well known for being the only coorbital system with comparable masses \citep{b15,b16,b17,b18,b19}.
As a consequence, they share the same mean orbit in horseshoe trajectories.
The discovery of a ring of particles also sharing the same orbital region makes the whole system even more complex.

%The information provided in the note of discovery is the location and the width of the ring \citep{b1}.
In the current work we investigate some basic features of this ring through analysis of Cassini images.
Is it a full ring or just a set of arcs? 
How wide is the ring?
What is its radial profile?
Then we numerically explore the lifetime of these particles, 
taking into account both gravity from Janus and Epimetheus and solar radiation pressure.
We find that the particles stay in the co-orbital region for just a few decades.
Since it is highly improbable that this is a temporary ring, we studied a physical model of particles production that could maintain 
the ring. The model is based on hypervelocity impacts of micrometeorites onto the surface of the co-orbital satellites.

The structure of this paper is as follows. 
The next section presents the photometric analysis of the images.
Section 3  describes our examination of the particles lifetime through a sample of numerical simulations.
The collisional model of particle production for the Janus Epimetheus ring are given in section 4.
In the last section we sumarize our conclusions.

\section{Photometric Analysis}

Our starting point to
 search for Cassini images that could have captured the Janus-Epimetheus ring
was the press release image   W1537029133 obtained 
in 2006 Sept. 15 (2006-258). 
 We have used the OPUS tool provided by the Ring-Moon system node of NASA's 
Planetary Data System to search images acquired in an interval of two days prior and 
two days after 2006-258 (ISS\_028RI\_HIPHASE001\_VIMS). We limited the results to data from both Wide Angle Camera 
(WAC) and Narrow Angle Camera (NAC) obtained through the clear filters (CL1 and CL2), 
which have a transmission band in the interval of 400 nm to 1000 nm for WAC, and 200 nm to 1000 nm
for NAC \citep{b3}. Using these parameters, 172 images (50 from NAC, and 122 from WAC) were retrieved.

A visual inspection was carried out to remove those images where the ring region would be out of the 
field of view or those where the ring region is saturated. After this process, the initial set was reduced 
to only 16 WAC images, all from the same day of PIA 08322, which are potentially useful for our purposes (Table 1). 
The best image of the ring obtained by the Cassini wide angle camera is shown in Fig. 1 (top panel).

At different geometries, with phase angles between zero and $150^\circ$  
images
$W1575913637$,
$W1524966389$,
$W1594709597$,
$W1622272726$,
$W1490835963$,
and
$W1615207585$), 
the search for images was made using 
the OPUS tool for the same camera (WAC), the same filters (CL1 and CL2),  and looking for regions that cover the location of the ring 
(nearby the orbits of Janus and Epimetheus).

\subsection{Calibration.}
The selected images were calibrated using the standard CISSCAL v.3 pipeline \citep{b3}, which provides an 
interactive tool to subtract instrumental backgrounds, to compensate the uneven bit weighting 
and optical distortions, as well as to correct the pixel-to-pixel relative sensitivity using the flatfield frames. 
We choose to convert the image count to normalized units of $I/F$, where $I$ is the intensity 
received from the ring and $\pi F$ is the solar flux at the distance of Saturn. The normalization is such 
that $I/F = 1$ for a perfectly reflecting Lambert surface illuminated  and viewed from normal incidence.

In some cases an additional step was applied to smooth the image, where we have used a moving box to replace the 
value of each pixel by the median of surrounding pixels. This procedure 
 removes cosmic rays and high-frequency noise in the image, making the ring easier to detect and quantify.
It was used when was necessary to get the radial profile.

\subsection{SPICE navigation.}
The images were navigated using the SPICE routines and the appropriate kernels for the given time interval 
of the observations. First we have computed the position of the spacecraft and the Sun 
relative to a body-fixed rectangular frame centered on Saturn (IAU$\_$SATURN), for which the reference plane 
coincides with the plane of the rings (assumed to be the equatorial). We were also able to determine 
the instrument field of view (FOV), as well as  the pointing vector and the ring plane intercept of each pixel. 

From the pointing vector we can determine the radial and longitudinal positions of every pixel in the 
ring plane, as well as the illumination angles, 
and the phase angle ($\alpha$).  
In order to check the accuracy of our code, we selected some high-resolution images, 
for which we could successfully determine the radial position of the narrow Keeler gap. 

Due to the geometry of the observation and the instrument characteristics, each image covers a wide range of co-rotating 
longitudes. To measure the azimuthal amplitude  we determine those pixels located at the radial distance 
of the ring and then we calculate the longitudinal position of the extremes of the arc in 
the IAU\_SATURN frame. When the longitudes are converted to a synodic system that rotates with the angular 
velocity of the satellites, we see that three images are enough to cover $320^\circ$  (Fig. 2), confirming the fact that 
the observed structure is indeed a complete ring. 
We do not detect any obvious longitudinal brightness variations in the ring.

\subsection{Radial profile.}
The ring is tenuous and it is 
in a region that also contains a broad dust population extending from the F to the G ring, and so
 it is not possible to safely distinguish the ring profile when taken across a single line of the image.
As an alternative, from image W1537029133 (top panel of Fig. 1) we extracted multiple radial scans and combined them to 
produce a better and smoother profile where the ring is clearly visible (bottom panel of Fig. 1). 

In the bottom panel of Fig. 1, the $I/F$ ratio is plotted as a function of the radial distance from Saturn.
Based on two local minima, we measured the width of the ring as approximately $7,500 \pm 500$~km, assuming an error up to one pixel.
That is approximately $50\%$ larger than previously announced \citep{b1}.

The estimated peak value of I/F  for the ring is $\sim 2.4\times10^{-5}$. This value is obtained by the 
difference between the ring peak and the background, which is assumed to be constant and equal to the value of the ring minimum at 
the inner edge (left red dot). It is worth  pointing out that all these quantities are consistent with two other images from 
where suitable radial profiles could be obtained by the same process of combination of multiple scans at different longitudes.

It can be seen that the ring presents an asymmetry, being brighter in the outward region. It is an unexpected feature, since 
the impact mechanism and the later evolution of the particles tend to spread the particles in such a manner to result a 
symmetrical profile whose peak is aligned with the source bodies. In our analysis, the peak of brightness  is located at 
$152359$~km, $\sim 920$~km outside the orbits of the satellites, and a 
possible explanation is that the outer region of the ring is more affected by light contamination due to the proximity with the G~ring. 
In the case of the background be considered as a straight line between the two minima, which is an extreme assumption, the $I/F$ value 
would be $1.9\times 10^{-5}$, i.e. about $20\%$ smaller. So, in this paper we will be considering values of $I/F$ that are 
at most $20\%$ larger than the case assuming an extreme background model. 
Still, if we suppose that the maximum of brightness of the ring matches the orbit of Janus and Epimetheus (blue line), 
and mirror the inner part of the ring to the right side, the ring maximum $I/F$ decreases to $2.3\times10^{-5}$ and 
its width is reduced by $700$~km, which is still  $36\%$ larger than previously thought.

\subsection{Photometry}
By calibrating and analyzing each image of the set of images that could have captured the J-E ring, 
we found that the ring was visible in images with very high phase angles (Sun-ring-Cassini), $175.4^\circ\le\alpha\le179.8^\circ$, 
indicating the strong forward scattering nature of the ring.
This behavior, along with the non-detection of the ring in observations of the same region made by Cassini at different 
geometries ($\alpha\sim24^\circ, \sim34^\circ, \sim65^\circ, \sim78^\circ, \sim97^\circ, \sim127^\circ$), 
indicate that Mie theory should be a useful approximation for a particle size distribution dominated by micrometric grains  \citep{b2}.
%From the clear filter transmittance  \citep{b3} we can place an upper limit of the observable particles size as 13 $\mu$m in radius, 
%that corresponds to approximately 20 times the incident wavelength \citep{b2}.
%Following \citet{b2} (section 13.42) and \citet{b2d} (Table 1 in p. 92-93).
Following \citet{b2} and \citet{b2d} we can place an upper limit of the  sizes of observable particles
as 20 times the incident wavelength , which  corresponds to approximately 13 $\mu$m in radius.
Therefore, despite of being a continuous full ring, it still can only be seen by Cassini's camera at very high phase angles.

For the images listed in Table 1, a visual inspection does not reveal any clump or gap along the ring azimuth, 
and the ring appears to be azimuthally uniform. The variation of the ring brightness depends only on the change of the phase angle.
In addition, the ring's absolute brightness depends on the emission angle of the pointing vector of each pixel, since incidence 
angle can be assumed  to be constant. In order to compare the I/F derived from the images with the theoretical model 
it is necessary to compensate for this variation, what can be accomplished by using the normal I/F given by
$(I/F)_{\perp} = (I/F) . \mu$, where $\mu$ is the cosine of the emission angle  (see section 4).

\section{Ring Lifetime}

We performed a set of tridimensional numerical simulations in order to study the orbital evolution of a
sample of  particles encompassing the orbits of 
 Janus and Epimetheus. These particles are influenced by the gravity of Saturn, including
the terms dues to $J_2$, $J_4$ and $J_6$, and the gravitational perturbations of Janus and Epimetheus. 
Since we are dealing with  $\mu$m sized particles, 
 in the numerical simulations were also included the force due to the solar radiation pressure  \citep{b4}.
This force can be given by the radiation pressure factor as  
 $$
 \beta=5.1\times 10^{-5} \frac{Q_{pr}}{\rho r} ,$$
 where $\rho$ is the particle density, $r$ its radius, and $Q_{pr}$ is the radiation pressure coefficient assumed to be the unit (ideal material) \citep{b22}.
 
The numerical integrations were made with the Burlish-Stoer integrator and
carried out using the Mercury package \citep{mercury}
with some modifications to include the solar radiation components \citep{b4}.
Saturn's orbit is sufficiently close to circular that we will assume a constant solar flux.
We did not consider the reflected sunlight from the planet
and the Yarkvovsky effect on the dust particles, since their effects are small enough to be  neglected  \citep{b4c}.
Tables 2 and 3 show the adopted values for  Saturn, Janus and Epimetheus.
We also considered the gravitational effects of the satellites 
Mimas, Tethys, Enceladus, Dione and Titan. The initial conditions 
of these satellites were generated from the
{\sl JPL Horizons System} at the same epoch (JD~2453005.5) as was used for Janus and Epimetheus.

The test  particles were equally distributed along 50 different values around the  initial
semi-major axis of  Epimetheus  (see Table 3), $a_{E} \pm 500$~km. 
The particles'  initial mean anomalies were placed between 0 and $360^{\circ}$ in steps of $1^{\circ}$.
The initial 
eccentricity and inclination of these particles were adopted to have the 
same values of the eccentricity and inclination of  Epimetheus,
since the values of these  two orbital elements are   larger than the eccentricity and 
inclination of Janus. 
The initial values of the argument of pericenter and longitude of the ascending node were also the same of those of Epimetheus.
The choice of such set of initial conditions was based on the idea of having particles that would have orbital behavior similar to the co-orbital
satellites.
%, which are expected to be more stable.
Each particle was numerically integrated for 100~yrs. Each time the distance 
between the particle and one of the satellites is smaller than the radius of the satellite
 (88 km for Janus and  58 km for Epimetheus), a 
collision is detected. The particle is assumed to escape the system when it encounters 
the outer edge of the F~ring (140,612 km) or the inner edge of the G~ring (166,000 km). All parameters
(time, position and velocity) were recorded during the numerical integrations.
The radius of the  particles ranges between 1 and $13\mu$m, each $1\mu$m. 
One set formed by $100\mu$m sized particles  was  also considered. Therefore, the ring 
is composed by 18,000 particles of each size.
Fig. 3 shows the lifetime of these $\mu$m sized  particles. 
Almost all of them collide with Janus or Epimetheus in less than 100~years. 
 Just a few hundred of them are removed by having encounters with the F or the G ring.

%The effects of the solar radiation pressure on the eccentricities of the larger particles ($> 10\mu$m)  are smaller when compared with the effects on the tiny particles. 
%Since these larger particles do not suffer a large variation in their eccentricities, they stay close to the orbits of the satellites, and are more rapidly removed  from the ring due to collisions.
 The larger particles are initially removed more rapidly by collisions (see Figure 3),
but note that this has only a modest effect on the average particle's lifetime (see Figure 4). 
% The average lifetime of the particles is generated by adding  the
% lifetime of $95\%$ of the particles (obtained from Fig. 3) 
% divided by $95\%$ of the particles. Table 4 gives the values used to 
% generate the particle's average lifetime (red line shown in Fig. 4). 
The relative velocities of the particles at the time they collide with one of the satellites (Janus or Epimetheus)
 are too low ($<0.09$km/s) to produce ejecta that can leave the satellite`s gravity \citep{b6}.

These calculations show that within only a few decades, the particles collide with Janus or Epimetheus,  which is 
consistent with the results of a previous 2-D study that did not take into account the effects 
of the solar radiation pressure  \citep{b5}. 
Due to the third dimension, in our more realistic model, the results show a fraction of particles surviving at least twice those of \citet{b5}. 
Therefore, the average lifetime of the ring particles is very short. The improbability that this is a temporary
ring implies that it must be continuously replenished.

\section{Particle Production}

Since the ring particles that collided with Janus or Epimetheus  
 had collision velocities that are at most on the order of just a hundred
meters per second, it is too low to eject material from the gravitational domain of the satellites 
\citep{b6}.
Therefore, another mechanism to supply material to the ring is needed.
Replenishment by ejecta produced by collisions between very energetic 
interplanetary dust particles and the host body has been
widely described in the literature for various targets, such as atmosphereless satellites  \citep{b6, b6b, b6c}, 
and the Pluto and Charon system  \citep{b8, b8b}. We developed a physical model to 
verify whether material removed from the coorbital satellites by interplanetary micrometeoroids impacts
would be consistent with the observed ring.   

Our physical model consists in a flux of interplanetary particles colliding with the satellites, 
generating ejecta \citep{b6,b7} that can supply material to replenish the  Janus and Epimetheus' ring. 
% From the literature  \citep{b6,b14} we estimated the averaged flux value that reaches the region of the satellites Janus and Epimetheus as being 
% ${ F }_{ imp }=1.678\times { 10 }^{ -16 }\textrm{g}\, {\textrm{cm}}^{-2} {\textrm{s}}^{-1}$, where was assumed that this flux is mainly composed by spherical particles of about 100 $\mu$m radius.
% It is important to emphasize that ${ F }_{ imp }$ is a function of the interplanetary flux (${ F }_{ \infty}$) and its velocity before reaching the planet (${ V}_{ \infty}$), and that the values of these quantities have large uncertainties
% (${ F }_{ \infty}$ is of one order of magnitude and ${ V}_{ \infty}$ is of the order of a factor two) \citep{b14, b20}.
 According to \citet{b20} the averaged flux value that reaches the region of the satellites Janus and Epimetheus is 
${ F }_{ imp }=3\times { 10 }^{ -18 }\textrm{g}\, {\textrm{cm}}^{-2} {\textrm{s}}^{-1}$, 
where was assumed that this flux is mainly composed by spherical particles of about 100 $\mu$m radius.

The amount of ejecta generated by the impacts on the surfaces of the satellites will depend on the dynamical 
conditions of the collisions and also on the physical characteristics of the impactor and the satellite.
 The rate of mass production can be given by \citep{b7}
$$
{\dot{ M}}={ F }_{ imp }YS\eta ,
$$
where  $Y$ is known as yield, $S$ is the cross section of the satellite and $\eta$ is given by
$
\eta =( { { V }_{ 0 } / { V }_{ esc } } )^{ \gamma  } ,
$
with $V_0$ being the minimum ejecta velocity,
$V_{esc}$  being the escape velocity, and $\gamma$ representing a slope, which assumes value equals 2,
 considering the satellites with pure ice surfaces \citep{b6}. 
 The value of the yield gives the ratio  between the ejecta mass and the impactor mass. 
 % Since the density of Janus and Epimetheus is low ($<0.8$g/cm$^3$), it is  reasonable to consider their surface as pure ice. 
  It is determined by experiments  and can be given by
$$
Y=2.642\times 10^{-5}{ m }_{ imp }^{ 0.23 }{ V }_{ imp }^{ 2.46 } ,
$$
where  $m_{imp}$ and  $V_{imp}$ are, respectively, the mass and the velocity of the impactor. 
According to the above references the value for the mass of the impactor is about $10^{-5}$ g. 
 Note that in this model the ejected material will have a density similar to the target surface (pure ice).
In Table 4 are presented the values of the rate of mass production, and other parameters, for each one of the satellites.

The steady-state mass resulting from such process is given by the product of the rate of the mass production 
by the average lifetime of the particles,
$
M={\dot{ M }}{ T}_{\rm{life}} .
$
 
The production rate of particles with mass between $m$ and $m_{max}$ can be inferred from \citep{b6,b7}
$$
\dot{ N } \left( >m \right) =\frac { 1-q  }{ q  } \frac {\dot{ M} }{ { m }_{ max } }
 \left( \frac { { m }_{ max } }{ m }  \right) ^{ q  } ,
$$
 where $q$ is the slope of a power law distribution.
 As can be seen in laboratory experiments, in most of the cases the value of this slope 
is 0.8   \citep{b6, b6c, b7}. 
 The quantity $m_{max}$ is the maximum particle mass of the 
ensemble that has the same size of the initial impactor, i.e, 100 $\mu$m.
The  number of particles for a given size range can be determined by multiplying $\dot{N}$  by the 
corresponding average lifetime, ${ T}_{\rm{life}}$, given in Table 5. 
In order to obtain the total size distribution, $N$, it is necessary 
 to add up the values found for each particle size range (every one micrometer).

  As the particle size increases, the production rate decreases according to a power law, while the average 
lifetime varies according to the particle size as shown in Fig. 4. 
The combination of these two parameters leads to the ring's steady state mass  \citep{b7, b10}. 
We found that the  normal optical depth,  ${\tau}$, of this steady state ring primarily arises
from the contribution of particles with radii between 1 and 4 $\mu$m (Fig. 4).

 \subsection{Converting Produced Particles Into a Measurable Quantity (I/F).}
In order to verify if the distribution of ring particles is compatible with the ring 
observed in Cassini images, we have to convert such distribution into a photometric quantity,  $I/F$.

As seen before, the images that show the ring were taken at high phase angles ($\alpha >175^{\circ}$).
Therefore, the ring is composed of small particles
 whose scattering properties can be well approximated with the Mie Theory.
For high phase angles,
the phenomenum  that 
 describes the observed light can be expressed through 
Chandrasekhar's equation \citep{b11}. Considering that the  ring is faint (normal optical depth, ${\tau} <<1$),
$I/F$ can be given by \citep{b12}
$$
 { I }/{ F } ={ { \varpi  }_{ 0 }P\left( \alpha  \right) {\tau}  }/{ 4\mu  } ,
$$
where $\mu$ is the  cosine of the emission angle, that was obtained from the images as being 
about $74.8^{\circ}$. $P\left( \alpha  \right)$ and ${ \varpi  }_{ 0 }$ are the phase 
function and the albedo of the particle  population, respectively. The former shows how the particles 
scatter radiation in the direction of the observer, whereas the latter (also known
as the reflection coefficient) is the ratio of reflected radiation from observed particles to 
incident radiation upon it. 
Both are computed by using a numerical code developed  for such purpose \citep{b13}. 
This code depends on the wavelength of scattered radiation (634 $nm$), the refraction index and the 
particle size distribution (found in the previous section). 
For an ice material without absorption, we have assumed that the real part of 
such index is 1.309 and its imaginary part is zero.

In Figure 5 is shown how the phase function changes according to the value of $\alpha$ for the 
range between $175^{\circ}$ and $180^{\circ}$. 
In order to obtain the value of $I/F$, we need to find the value 
of ${\tau}$. The normal optical depth, ${\tau}$, is given by the ratio between two areas:
$$
{\tau} = {A_{cs}}/{{ A }_{ ring }},
$$
where $A_{cs}$ is the cross section area resulting from all the visible particles that form
the ring, and  ${ A }_{ ring }$ is the area in Saturn's equatorial plane that contains the ring.
  Assuming a triangular profile for the ring's brightness, we can approximate
$$
{ A }_{ ring }\quad =\quad( \pi/2) \left[ { \left( { R }_{ a }+{ L }/{ 2 } \right)  }^{ 2 }-{ \left( { R }_{ a }-{ L }/{ 2 } 
\right)  }^{ 2 } \right]=3.59 \times 10^{9} {\rm km}^2,$$  
where $R_a$ is the distance between the center of planet and the ring ($152,359.6$ km) and $L$ is
the width of the ring ($7,500$ km). 
 The factor 2 in the above expression arises from the assumed shape of the profile.

The $A_{cs}$ is computed through the addition of the cross section areas of each particle of the ring. 
In order to do that, we divided the set of particles in 13 particle's size ranges of 1 $\mu m$. Then,  
for each range, we obtained the average cross section area, the average lifetime 
($\bar { T^i }_{life}$) and the difference of the rate production of range extremes ($\Delta \dot{N}^i_T$).
By multiplying these three values, and summing up the results for all ranges, we obtain
$$
 A_{cs} =  \sum _{ i = 1\leftrightarrow 2 }^{ 12\leftrightarrow 13 }{\bar{A}^i \Delta {\dot{N}_T^i}}
 {\bar{T}_{life}^i} = 1.23\times 10^{4} {\rm km}^{2}  \quad .
$$
Thus, the total normal optical depth is given by
$$
 {\tau} =  \sum _{ i = 1\leftrightarrow 2 }^{ 12\leftrightarrow 13 }{{{\tau}}^i}  \quad ,
$$
where ${{{\tau}}^i}$ is the  normal optical depth due to each range of particles size. The values
of $\bar{A}^i$ and ${{\tau}}^i$ are given in Table 5. 
Therefore, if ${ F }_{ imp }=3\times { 10 }^{ -18 }\textrm{g}\, {\textrm{cm}}^{-2} {\textrm{s}}^{-1}$,
then the total normal optical depth of the visible particle ring should be about  $6\times 10^{-8}$.

Therefore, for any given value of $\alpha$ one can compute the $I/F$ from the current model,  where the albedo was always assumed to be equal one.
For example, for $\alpha=175.8^{\circ}$ the corresponding value of the phase function is about 70 (Fig. 5).
A comparison of the results of our model with the values
 obtained by photometry from a Cassini image is presented in Figure 6.
It shows a curve (full black line) of the value of the normal $I/F$
($(I/F)_{\perp} = (I/F) . \mu$) as a function of the phase angle ($\alpha$) for our physical model.
Note the large decrease in the value of $(I/F)_{\perp}$ as the phase angle reaches values only two or three 
degrees smaller than $180^\circ$. The blue dots in Figure 6 indicate photometric data obtained from the same
Cassini image shown in Figure 1. 

From the best ring image, considering 
its background level and the camera sensitivity, we can estimate that the ring is only visible by the Cassini spacecraft
when it forms a phase angle (Sun-ring-Cassini) in the range from $140^\circ$ to $180^\circ$. 
That has also been confirmed by the non-detection of the ring in observations of the same region made by Cassini at different 
geometries (red triangles).
Thus, for the Cassini spacecraft orbiting around Saturn and pointing its camera towards the ring, the ring will ``not be ��visible"�� most of the time (Fig. 6).
Therefore, we say that the ring behaves like a firefly. The analogy is that it can only be viewed by the spacecraft from time to time, 
when Cassini, the ring and the Sun are arranged in a particular geometric configuration. 
Otherwise, it remains ``in the dark", not visible to Cassini's cameras. 
Although a wider range of particles, in terms of size and quantity, might be found in the ring, only those particles 
in the range of 1 to 4 $\mu$m are responsible for almost all of the light captured by the Cassini spacecraft camera (Fig. 4).

The physical model results (full black line in Fig. 6) were generated with a micrometeoroids flux of 
$3\times 10^{-18} {\rm g}/{\rm cm}^{2} {\rm s}$, 
which is  the most up-to-date value found in the literature  \citep{b20}.
From the zoom of Figure 6 we note that the values of the photometric data (blue dots) are not very well fitted by the physical model (full black line). However, this discrepancy is  acceptable taking into account the  uncertainties of the parameters involved.
 For instance, if we assume a flux one order of magnitude higher (dashed line), which is perfectly plausible, the physical model fits much closely the observational data (zoom of Figure 6).
Consequently, this ring with very short lifetime particles can be accordingly replenished
by particles generated through interplanetary micrometeoroid collisions with the surfaces of Janus and Epimetheus.

\section{CONCLUSIONS}

In this work we performed a set of studies on the Janus-Epimetheus ring.
From photometric analysis of Cassini images we found that in most of the images, 
the ring extends along a wide longitude range, and a combination of just three of 
these images shows that the longitudes covered by the ring make an almost complete ring, 
not just a series of arcs.
Furthermore, no significant variation in brightness with longitude was 
visually found. Thus, we conclude that this ring is continuous and smooth. 
The width of the ring is about $50\%$ larger than previously announced \citep{b1}.

%The low peak value of I/F indicates that the ring is the faintest planetary 
%ring ever observed. 

%Despite of being a continuous full ring, it could only be seen by Cassini's camera at very high phase angles,
%indicating the ring is mainly composed of microsized particles.

In our search for Cassini images that could have observed the ring, we verified that it appeared only in high phase angle images.
What indicated the ring is mainly composed of microsized particles.

Then, we studied the lifetime of the ring particles.
The results showed that within only a few decades, the particles collide with Janus or Epimetheus.
Therefore, the lifetime of the  ring particles is too short, suggesting the ring must be continuously replenished. 

In order to provide such replenishment we considered a collisional model of micrometeorites on the surface of the two satellites.
As the particle size increases, the production rate decreases according to a power law, 
 while the average lifetime is roughly constant around 15 years.
The combination of these two parameters leads to the ring's steady state mass. 
We found that the optical depth of this steady state ring primarily arises
from the contribution of particles with radii between 1 and 4 $\mu$m.
The outcome of this model of particles production showed to be compatible with the Janus-Epimetheus ring data obtained from image analysis.
The ring has also been identified at lower phase angles ($\sim146^\circ$  and $\sim157^\circ$) with normal $I/F$ at least one order of magnitude 
lower than those presented in this work (Matthew Hedman, private communication). 
That is in full agreement with our results and conclusions.

\floattable
\begin{deluxetable}{ccCrlc}
\tablecaption{Summary of the useful images retrieved from PDS Ring Node \label{tab:images}}
\tablecolumns{6}
\tablenum{1}
\tablewidth{0pt}
\tablehead{
\colhead{Image} &
\colhead{time} &
\colhead{Exposure time} & \colhead{$\Delta \theta$} & \colhead{Phase angle} \\
\colhead{} & \colhead{(day 258)} &
\colhead{(ms)} & \colhead{($^\circ$)} & \colhead{($^\circ$)}
}
\startdata
W1537006439 & 09:42:15.769 & 80 & 159 & 177.2 -- 179.5 \\
W1537006505 & 09:43:21.805 & 15 & 159 & 177.2 -- 175.5  \\
W1537007792 & 10:04:48.760 & 80 & 119 & 177.1 -- 179.1  \\
W1537007825 & 10:05:21.614 & 380 & 119 & 177.1 -- 179.1 \\
W1537008000 & 10:08:16.793 & 50 & 119 & 177.1 -- 179.1  \\
W1537009120 & 10:26:56.752 & 80 & 119 & 177.1 -- 179.1  \\
W1537011801 & 11:11:37.734 & 80 & 67 & 176.9 -- 177.7  \\
W1537011834 & 11:12:10.588 & 380 & 67 & 176.9 -- 179.2 \\
W1537028326 & 15:47:02.613 & 80 & 144 & 176.3 -- 179.2  \\
W1537028414 & 15:48:30.654 & 20 & 144 & 176.3 -- 179.2  \\
W1537028438 & 15:48:53.068 & 3200 & 144 & 176.3 -- 179.2 \\
W1537028473 & 15:49:29.080 & 1200 & 144 & 176.3 -- 176.2 \\
W1537028702 & 15:53:18.611 & 80 & 49 & 176.6 -- 179.5 \\
W1537029078 & 15:59:34.608 & 80 & 145 & 175.5 -- 178.4  \\
\textbf{W1537029133} & 16:00:29.581 & 150 & 145 & 175.5 -- 178.4 \\
W1537029166 & 16:01:02.650 & 20 & 145 & 175.5 -- 178.4 \\
W1537029190 & 16:01:25.063 & 3200 & 145 & 175.5 -- 178.4 \\
\enddata
\tablecomments{The time corresponds to the MID\_TIME of each frame
and $\Delta\theta$ is the azimuthal coverage of the ring.
The image W1537029133 was identified as PIA 08322.}
\end{deluxetable}

\floattable
\begin{deluxetable}{ccCrlc}
\tablecaption{Saturn physical data$^a$ \label{tab:saturn}}
\tablecolumns{2}
\tablenum{2}
\tablewidth{0pt}
\tablehead{
\colhead{parameter} &
\colhead{value} &
%\colhead{Seeing} & \colhead{Filter} & \colhead{Inst.} \\
%\colhead{(YYYY-mm-dd)} & \colhead{(d)} &
%\colhead{(arcsec)} & \colhead{} & \colhead{}
}
\startdata
$m_{S}$~(g) & $5.69 \times 10^{29}$  \\
radius~(km) &  60330 \\
$J_2$ & $1.6298 \times 10^{-2}$\\
$J_4$ & $-9.15 \times 10^{-4}$\\
$J_6$ & $1.03 \times 10^{-4}$\\
\enddata
\tablecomments{ ($a$) \citet{b25}.}
\end{deluxetable}

\floattable
\begin{deluxetable}{ccCrlc}
\tablecaption{Janus and Epimetheus data \label{tab:janus}}
\tablecolumns{3}
\tablenum{3}
\tablewidth{0pt}
\tablehead{
\colhead{} &
\colhead{Janus} &
\colhead{Epimetheus} 
}
\startdata
$m^b$ &  $5.83 \times 10^{-9}~m_{S}$ &  $1.17 \times 10^{-9}~m_{S}$\\
$a^c$(km) &  151440 & 151490 \\
$e^c$ & 0.0068 &  0.0097 \\
$I^c$ (deg) & 0.16 & 0.35\\
$\varpi^c$ (deg) & 287.60 & 41.11\\
$\Omega^c$ (deg) & 47.87 & 82.60\\
$\lambda^c$ (deg) & 90.59 & 278.15\\
\enddata
\tablecomments{$m_{S}$ and $m$ are the masses of Saturn and the satellites, respectively. 
$a$, $e$, $I$, $\varpi$, $\Omega$ and $\lambda$  are the 
semimajor axis,   eccentricity, inclination, longitude of the pericenter, 
longitude of the ascending node and the mean longitude, respectively.
($b$) \citet{b26}, ($c$) \citet{b27}.}
\end{deluxetable}

\floattable
\begin{deluxetable}{ccCcccc}
\tablecaption{Rate of mass production \label{tab:production}}
\tablecolumns{7}
\tablenum{4}
\tablewidth{0pt}
\tablehead{
\colhead{Satellite} &
\colhead{$S(km^2)$} &
\colhead{$V_0 (m/s)$} & 
\colhead{$V_{esc} (m/s)$} & 
\colhead{$V_{imp} (km/s)$} &
\colhead{$\eta$} & 
\colhead{$\dot{M} (kg/s)$}
}
\startdata
         Janus & $ 2.52 \times 10^{4} $ & $28.5$ & $53.0$ & $23.41$ & $0.287$ & $5.85\times 10^{-3}$ \\ 
Epimetheus & $ 1.01 \times 10^{4} $ & $28.5$ & $35.0$ & $23.41$ & $0.654$ & $5.35\times 10^{-3}$ \\ 
\enddata
%\tablenotetext{a}{At exposure start.}
\tablecomments{Cross section ($S$), the minimum ejection velocity ($V_0$), the escape velocity ($V_{esc}$), 
the impact velocity ($V_{imp}$),  the parameter $\eta$ and the mass production rate ($\dot M$).}
\end{deluxetable}

\floattable
\begin{deluxetable}{ccCrlc}
\tablecaption{Mean Lifetime ($T^i_{life}$) and steady-state mass (M) \label{tab:state}}
\tablecolumns{6}
\tablenum{5}
\tablewidth{0pt}
\tablehead{
\colhead{$r_{min}$($\mu m$)} &
\colhead{$r_{max}$($\mu m$)} &
\colhead{$T^i_{life}$(years)} & \colhead{M(kg)} & \colhead{$\bar{A}^{i}$($\mu m^2$)} & \colhead{${{\tau}^i}$}
}
\startdata
$1$ & $2$ & $16.67$ & $3.42\times 10^{5}$ & $5.63$ & $2.48\times 10^{-8}$\\ 
$2$ & $3$ & $15.67$ & $1.82\times 10^{5}$ & $18.14$ & $1.09\times 10^{-8}$\\ 
$3$ & $4$ & $15.33$ & $1.40\times 10^{5}$ & $39.98$ & $6.58\times 10^{-9}$\\ 
$4$ & $5$ & $14.81$ & $1.17\times 10^{5}$ & $62.10$ & $4.47\times 10^{-9}$\\ 
$5$ & $6$ & $14.66$ & $1.05\times 10^{5}$ & $93.51$ & $3.33\times 10^{-9}$\\ 
$6$ & $7$ & $14.86$ & $9.78\times 10^{4}$ & $131.22$ & $2.67\times 10^{-9}$\\ 
$7$ & $8$ & $14.43$ & $8.87\times 10^{4}$ & $175.20$ & $2.12\times 10^{-9}$\\ 
$8$ & $9$ & $13.97$ & $8.15\times 10^{4}$ & $225.48$ & $1.72\times 10^{-9}$\\ 
$9$ & $10$ & $13.74$ & $7.63\times 10^{4}$ & $282.03$ & $1.45\times 10^{-9}$\\ 
$10$ & $11$ & $13.44$ & $7.15\times 10^{4}$ & $344.83$ & $1.23\times 10^{-9}$\\ 
$11$ & $12$ & $13.18$ & $6.76\times 10^{4}$ & $413.95$ & $1.06\times 10^{-9}$\\ 
$12$ & $13$ & $12.40$ & $6.13\times 10^{4}$ & $489.37$ & $8.87\times 10^{-10}$\\ \enddata
%\tablenotetext{a}{At exposure start.}
\tablecomments{Average cross section area ($\bar{A}_{i}$) and normal optical depth (${{\tau}^i}$) for each one of the size ranges between 1 and 13 $\mu m$.
The steady state mass is computed assuming a nominal impact flux of $3\times { 10 }^{ -18 }\textrm{g}\, {\textrm{cm}}^{-2} {\textrm{s}}^{-1}$.}
\end{deluxetable}

\floattable
\begin{deluxetable}{cccc}
\tablecaption{Pixels coordinates from image W1537029133 \label{tab:coordinates}}
\tablecolumns{5}
\tablenum{6}
\tablewidth{0pt}
\tablehead{
\colhead{Coordinate} & \colhead{Coordinate} & \colhead{Phase Angle} & \colhead{$(I/F)_{\perp}$}  \\
\colhead{$X$}             & \colhead{$Y$}           & \colhead{($^\circ$)} & \colhead{}
}
\startdata
103 & 264 & 175.49 & 7.8022e-06 \\
103 & 259 & 175.50 & 8.3329e-06 \\
104 & 251 & 175.52 & 9.6913e-06 \\
126 & 219 & 175.70 & 1.3674e-05 \\
139 & 209 & 175.80 & 1.4769e-05 \\
141 & 208 & 175.81 & 1.4515e-05 \\
156 & 198 & 175.92 & 1.5633e-05 \\
\enddata
\tablecomments{The origin of the coordinate system (0,0) is at the lower left corner of the image.}
\end{deluxetable}

\begin{figure}
\gridline{
\includegraphics[width=12cm,angle=90]{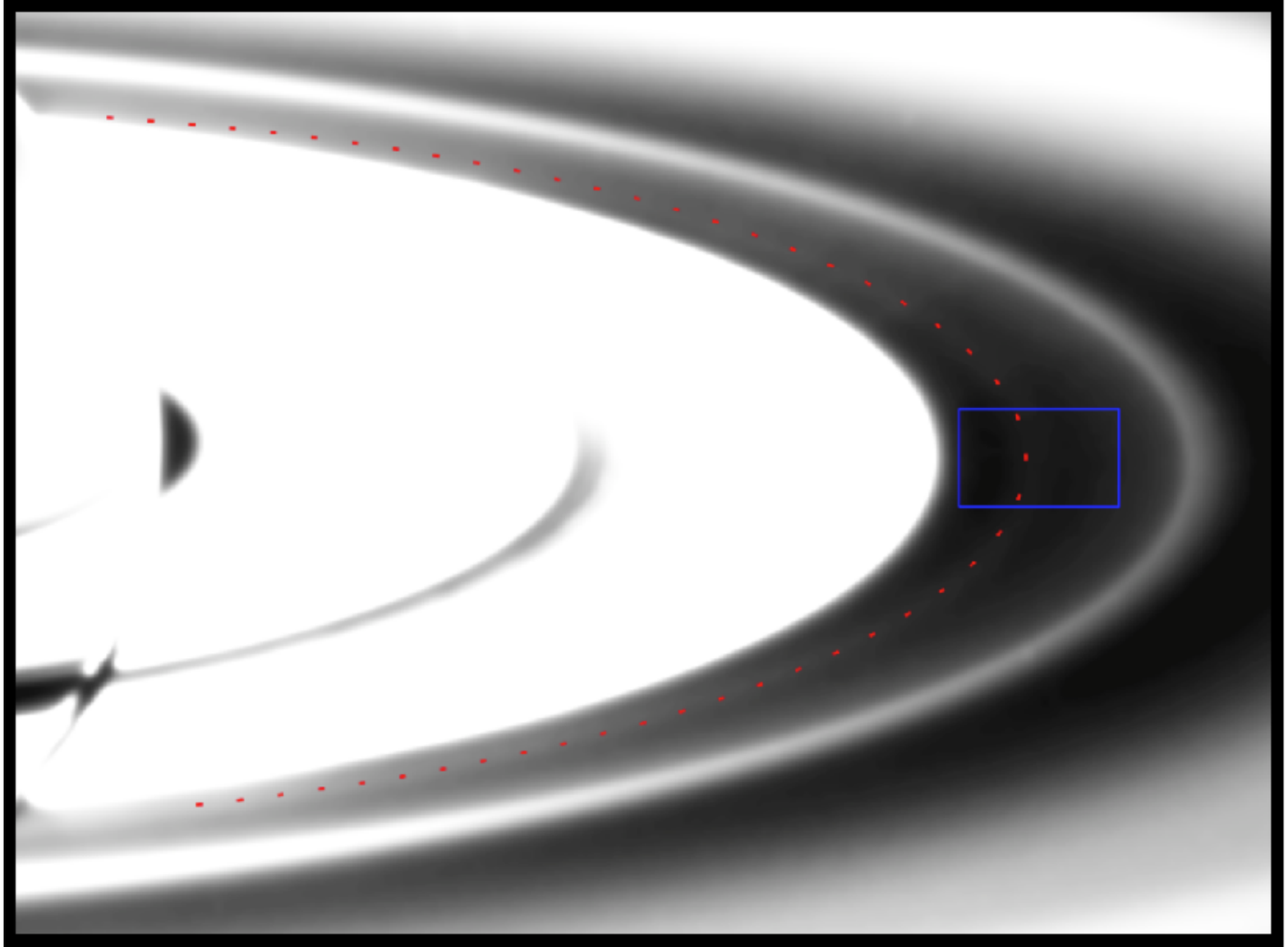}
         }
         \gridline{
\includegraphics[width=12cm,angle=0]{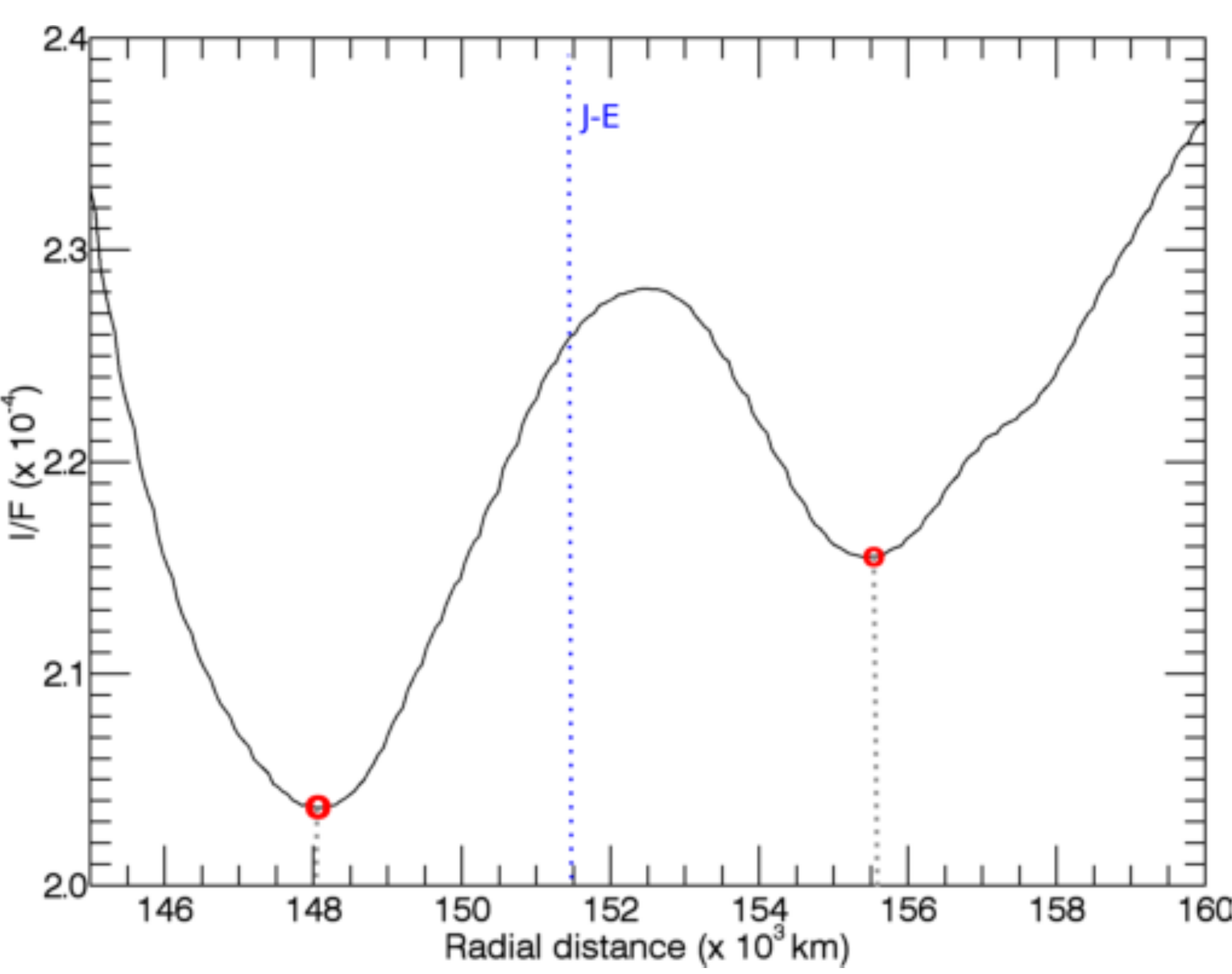}
         }
\caption{
       Saturn's rings image  (W1537029133) from Cassini spacecraft camera (top panel). 
The location of the Janus-Epimetheus ring is indicated by a red dashed line. 
A plot of the I/F for the region indicated by the small blue box is presented in the bottom panel.
This plot also shows the width of the ring ($\approx 7,500$ km) and the location of averaged 
orbits of the co-orbital moons, Janus and Epimetheus. 
Note an asymmetry, the peak does not coincide with the mean orbital radius of the co-orbital moons.
      }
%\label{Fig1}
\end{figure} 
\clearpage

\begin{figure}
\begin{center}
\includegraphics[width=12cm,angle=90]{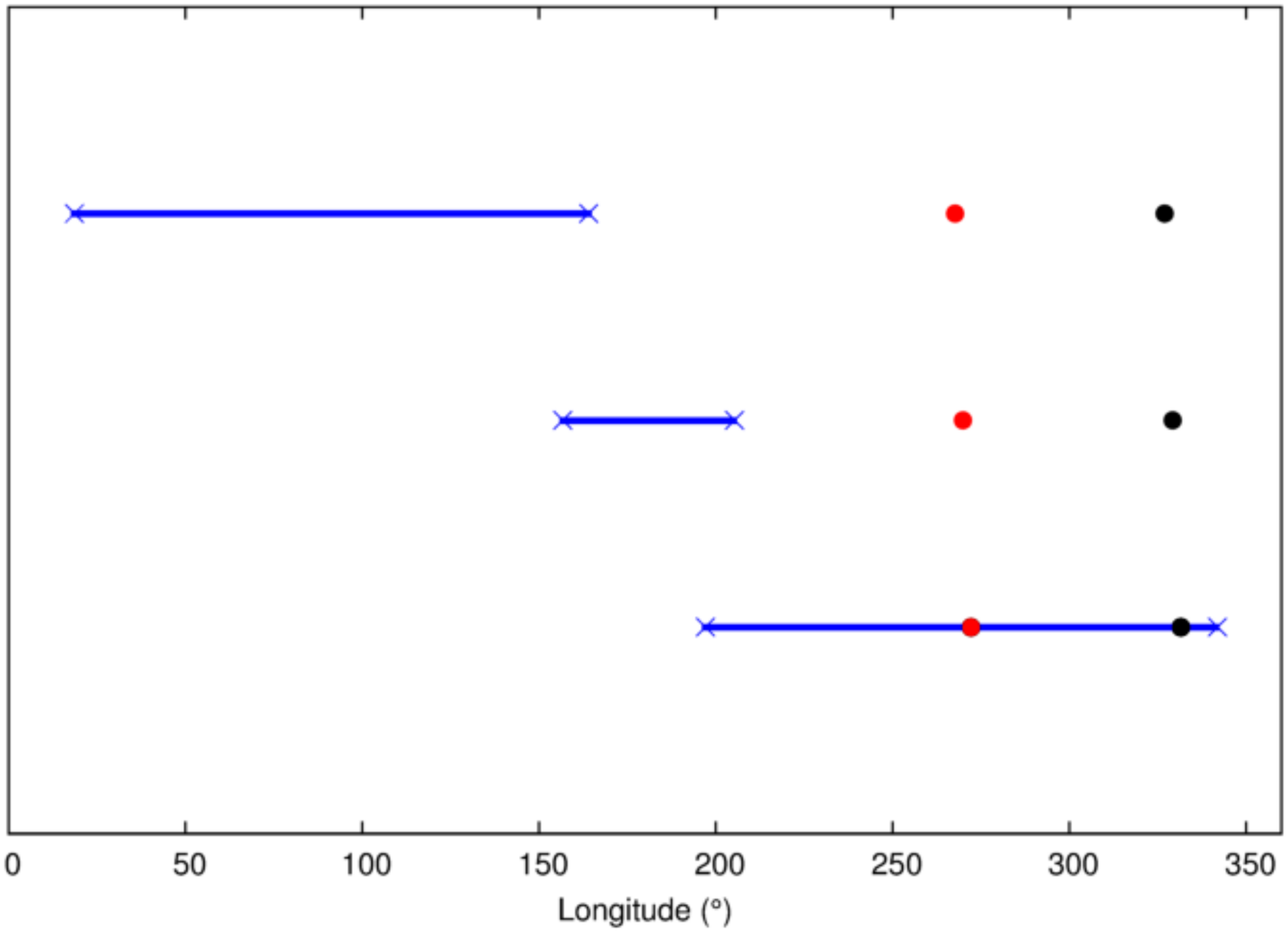}
\end{center}
\caption{ Longitudinal coverage of the ring for three different images measured in a frame that 
rotates with angular velocity equal to the mean motion of the satellites. For each image the positions of Janus 
and Epimetheus are represented by red and black dots, respectively. 
Note that the zero point of this longitude is arbitrarily set for display purposes.
From top to bottom, the data correspond to images W1537028326, W1537028702 and W1537029078.}
%\label{Fig2}
\end{figure}
\clearpage

%\begin{figure}
%\begin{center}
%\includegraphics[width=12cm]{figures/profile2b}
%\end{center}
%\caption{ Radial profile (black line) obtained by the combination of 50 radial scans from image W1537029133. The blue dashed line indicates 
%the mean semi-major axis of Janus and Epimetheus, while the other vertical dotted lines indicate the ring edges (red marks). 
%A constant background equal to the ring minimum 
%is shown by the red line. %, and the green line denotes a simple background model. 
%The grey line is a reflection of the ring profile interior to the orbit of the satellites.  }
%\end{figure}
%\clearpage

\begin{figure}
\begin{center}
\includegraphics[height=10cm]{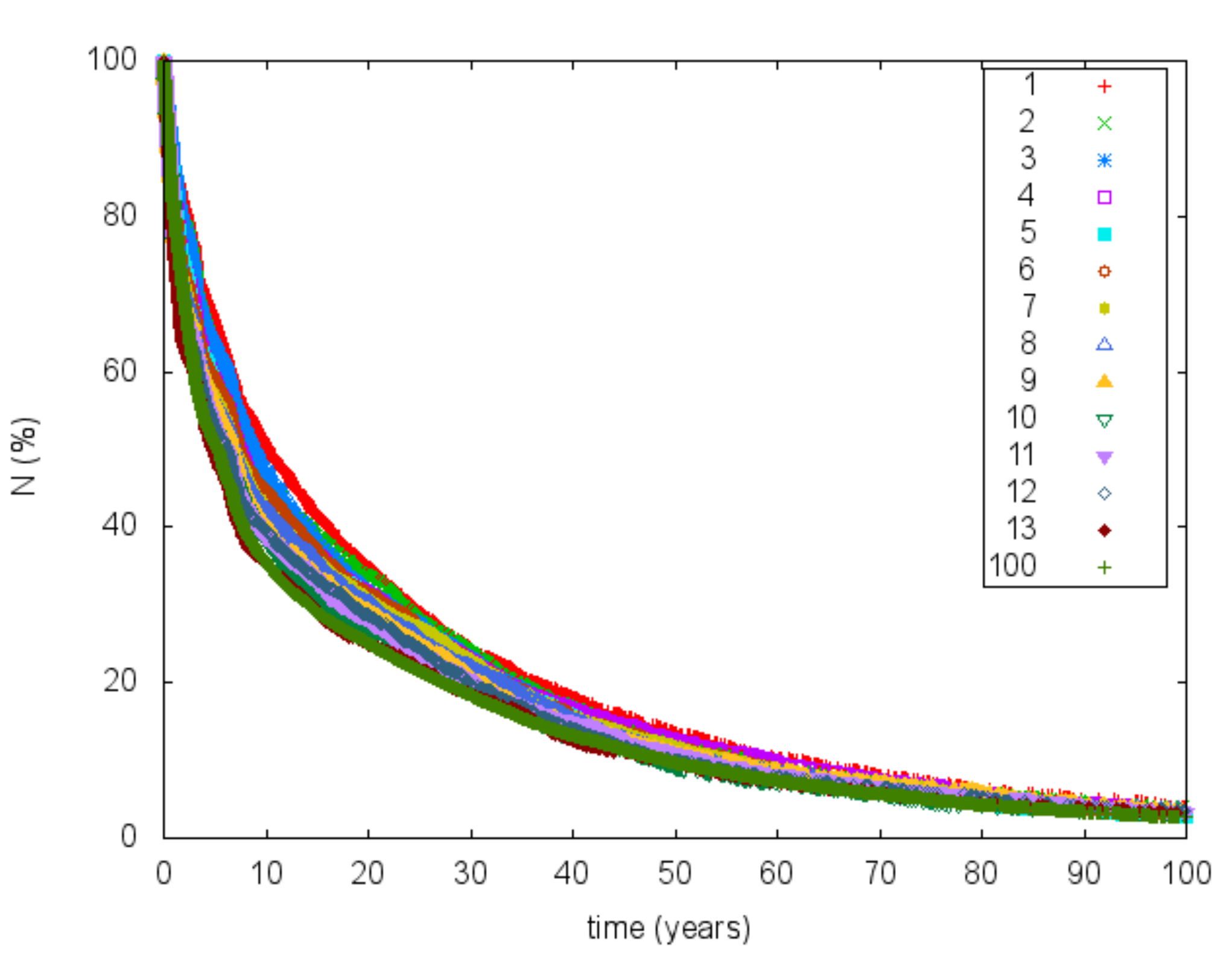}
%}
\end{center}
\caption{ Temporal evolution of the remaining amount of particles ($N$) for different sizes. 
Each color or symbol indicates the size of the particle,
from 1 to $13\mu$m, each $1\mu$m, and $100\mu$m.}
%\label{Fig3}
\end{figure}
\clearpage

%\begin{figure}
%\centering
%\begin{center}
%\includegraphics[height=3.8cm]{figures/n50.pdf}
%\includegraphics[height=8cm]{figures/meio}
%\includegraphics[height=3.8cm]{figures/p50.pdf}
%\end{center}
%\caption{ Relative velocities of the particles at the time they collide with 
%Janus or Epimetheus.} 
%\end{figure} 
%\clearpage

\begin{figure}
\begin{center}
\includegraphics[width=12cm, angle=-90]{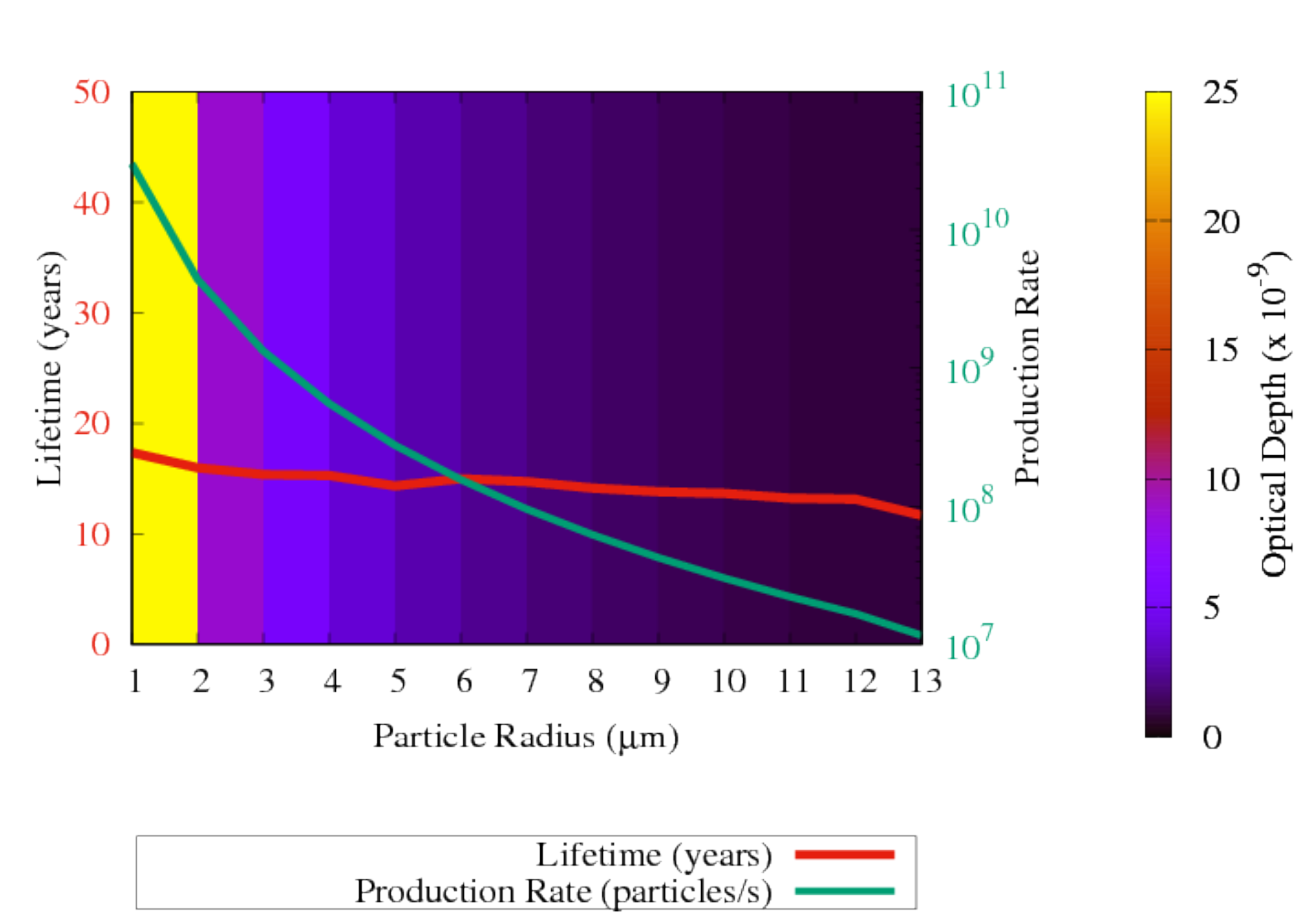}
\end{center}
\caption{
      Normal optical depth values computed for each particle size range of the physical model (color coded strips).
The normal optical depth depends on the particle's production rate (green line), that decreases according 
to a power law, and the particle's average lifetime (red line).
The production rate is computed assuming a nominal impact flux of $3\times { 10 }^{ -18 }\textrm{g}\, {\textrm{cm}}^{-2} {\textrm{s}}^{-1}$.
Note that the optical depth of this ring is dominated by the contribution of particles with radii between 1 and 4 $\mu$m. 
The average lifetime was obtained by the sum of the lifetime for each particle divided by the number of particles.
      }
%\label{Fig4}
\end{figure} 
\clearpage

\begin{figure}
\begin{center}
\includegraphics[width=10cm, angle=-90]{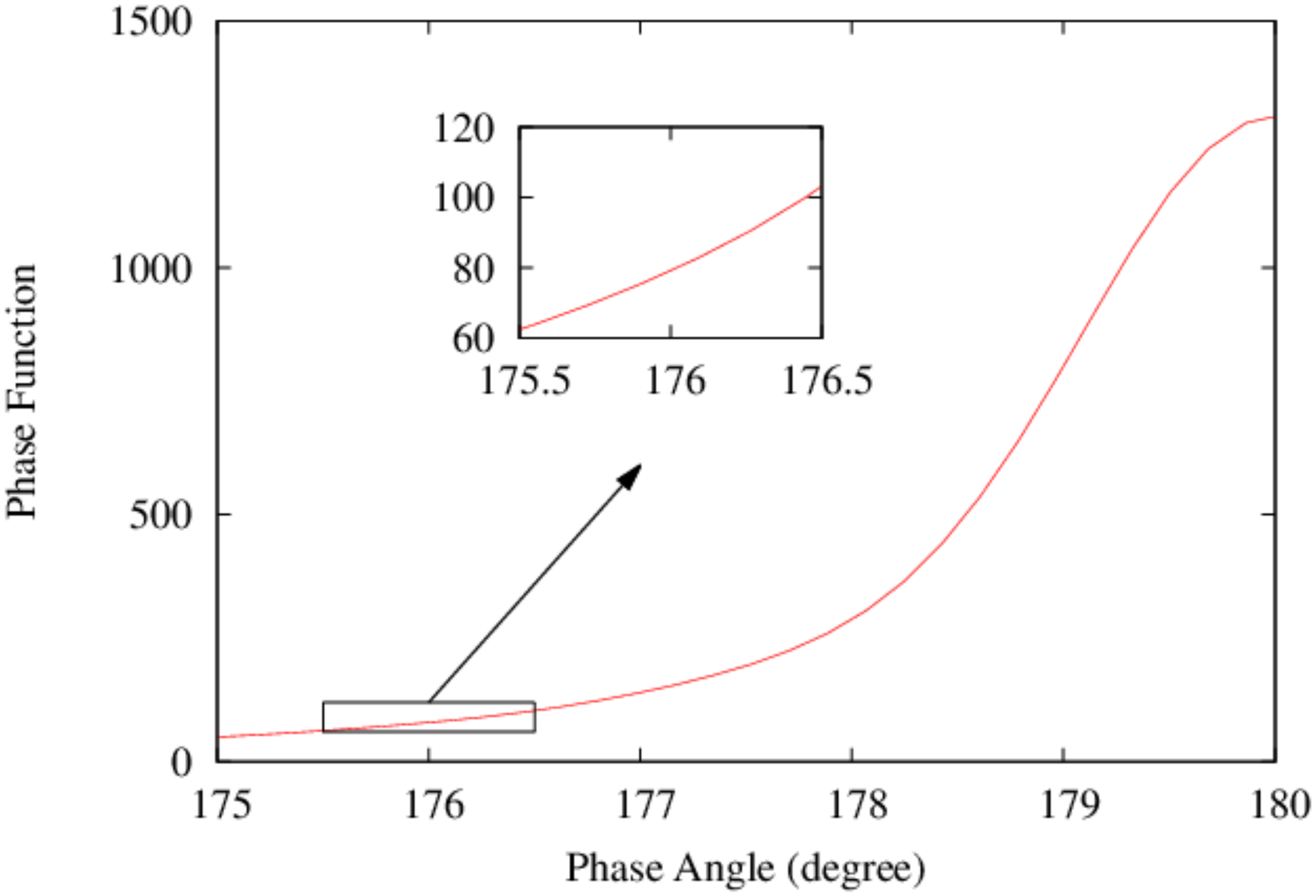}
\end{center}
\caption{ Phase function versus phase angle.  
 This is a phase curve for the particle size distribution given in Table 5.
It was computed by using a numerical code developed  for such purpose \citep{b13}. 
The insertion shows a zoom illustrating how fast the phase function changes around $176^\circ$, 
which is nearby where the image data were obtained.
}
%\label{Fig5}
\end{figure}

\begin{figure}
\begin{center}
\includegraphics[width=14cm]{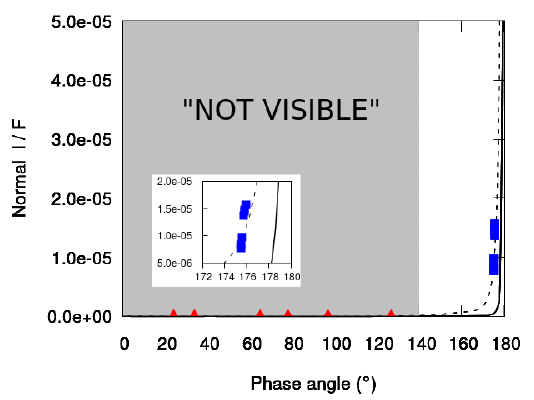}
\end{center}
\caption{
       The visibility range: plot of $(I/F)_{\perp}$ as a function of the phase angle ($\alpha$). 
        The full black line was derived from the physical model assuming a nominal impact flux 
       of $3\times { 10 }^{ -18 }\textrm{g}\, {\textrm{cm}}^{-2} {\textrm{s}}^{-1}$, while the dashed line 
       was derived assuming a flux one order of magnitude higher.
%       adopting
       %$v_{\infty}=9.5{\rm km/s}$ and
%       $F_{\infty}=2.6\times 10^{-19} {\rm g}/{\rm cm}^{2} {\rm s}$.
       Note the abrupt change in the values of $(I/F)_{\perp}$ as $\alpha$ gets close to $180^\circ$.
 The blue squares are data extracted from the same Cassini image shown in Fig. 1 and their coordinates are given in Tab. 6.
The red triangles indicate $\alpha$ of images where the ring could not be identified.
For $0^\circ\le\alpha\le140^\circ$ we have that $(I/F)_{\perp}\approx 0$ (gray zone).
Therefore, despite of the J-E ring be a continuous full ring, it was only be seen
by Cassini's camera within a short arc of phase angle ($>140^\circ$).
The insertion is a zoom to show how well the data fits the physical model.
}      
%\label{Fig6}
\end{figure} 
\clearpage

\acknowledgments
The Cassini image data used in this work was provided by NASA's Astrophysics Data System.
This work, which was supported by  FAPESP (proc. 2011/08171-3, proc. 2016/24561-0), CNPq and CAPES, 
was improved by helpful discussions with Nilton do Rosario, Doug Hamilton, Roberto Vieira Martins and Matthew Hedman.
The authors would like to thank a generous referee that contributed significantly for this final version of the paper.



\end{document}